\documentclass[
 reprint,%
aip, jap,
superscriptaddress,
floatfix
]{revtex4-1}

\usepackage{makeidx}

\usepackage{graphicx}
\usepackage{amsmath}
\usepackage{epstopdf}
\usepackage{textcomp}
\usepackage{epstopdf}
\usepackage{tabularx}
\usepackage{ulem} 
\usepackage{color}

\usepackage{natbib}
\usepackage{bm}%
\usepackage[colorlinks=true,linkcolor=blue]{hyperref}%
\expandafter\ifx\csname package@font\endcsname\relax\else
 \expandafter\expandafter
 \expandafter\usepackage
 \expandafter\expandafter
 \expandafter{\csname package@font\endcsname}%
\fi
\makeindex

\begin{document}

\title{\color{blue}Optimizing the spin-pumping induced inverse spin Hall voltage by crystal growth  in Fe/Pt bilayers}

\author{Evangelos Th. Papaioannou}

\thanks{Author to whom correspondence should be addressed. Email: epapa@physik.uni-kl.de}
\affiliation{Fachbereich Physik and Landesforschungszentrum OPTIMAS, Technische Universit\"{a}t Kaiserslautern,
Erwin-Schr\"{o}dinger-Str. 56, 67663 Kaiserslautern, Germany}

\author{Philipp Fuhrmann}
\affiliation{Fachbereich Physik and Landesforschungszentrum OPTIMAS, Technische Universit\"{a}t Kaiserslautern,
Erwin-Schr\"{o}dinger-Str. 56, 67663 Kaiserslautern, Germany}

\author{Matthias B. Jungfleisch}
\affiliation{Fachbereich Physik and Landesforschungszentrum OPTIMAS, Technische Universit\"{a}t Kaiserslautern,
Erwin-Schr\"{o}dinger-Str. 56, 67663 Kaiserslautern, Germany}

\author{Thomas Br\"{a}cher}
\affiliation{Fachbereich Physik and Landesforschungszentrum OPTIMAS, Technische Universit\"{a}t Kaiserslautern,
Erwin-Schr\"{o}dinger-Str. 56, 67663 Kaiserslautern, Germany}

\author{Philipp Pirro}
\affiliation{Fachbereich Physik and Landesforschungszentrum OPTIMAS, Technische Universit\"{a}t Kaiserslautern,
Erwin-Schr\"{o}dinger-Str. 56, 67663 Kaiserslautern, Germany}

\author{Viktor Lauer}
\affiliation{Fachbereich Physik and Landesforschungszentrum OPTIMAS, Technische Universit\"{a}t Kaiserslautern,
Erwin-Schr\"{o}dinger-Str. 56, 67663 Kaiserslautern, Germany}

\author{J\"{o}rg L\"{o}sch}

\affiliation{Institut f\"{u}r Oberfl\"{a}chen- und Schichtanalytik (IFOS) and Landesforschungszentrum OPTIMAS, Trippstadter Str. 120, 67663 Kaiserslautern, Germany} 

\author{Burkard Hillebrands}
\affiliation{Fachbereich Physik and Landesforschungszentrum OPTIMAS, Technische Universit\"{a}t Kaiserslautern,
Erwin-Schr\"{o}dinger-Str. 56, 67663 Kaiserslautern, Germany}


\begin{abstract}

We examine the influence of crystal growth on the spin-pumping induced inverse spin Hall effect in Fe/Pt bilayers. The morphology of the Fe/Pt interface influences the effective spin mixing conductance. The increase of growth temperature leads to smoother and larger grains at the interface that enhance the effective spin mixing conductance. The spin current injection efficiency into Pt, measured by the inverse spin Hall effect, is maximized by optimizing the epitaxy of Pt on Fe. In magnetic field dependent measurements, the presence of a strong magnetic anisotropy gives rise to two distinct inverse spin Hall effect voltage peaks.

\end{abstract}

\pacs{}

\keywords{}

\maketitle


\noindent Magnon spintronics -- an emerging sub-field of spintronics in which magnons, the quasi-particles of the magnetization dynamics, are used as information carriers --  has attracted considerable attention in the last years. \cite{saitoh:182509,Kajiwara} A powerful method to detect magnons is the combination of spin pumping and the inverse spin Hall effect. The spin-pumping effect allows for the injection of a spin current from a ferromagnetic (FM) layer into an attached non-magnetic metal (NM) layer. \cite{Tserkovnyak} This spin current is subsequently transformed into a charge current by the inverse spin Hall effect (ISHE). \cite{Hirsch,saitoh:182509} Many aspects of magnetization dynamics in FM/NM bilayers have been investigated by means of spin pumping and ISHE in various kinds of materials. \cite{Goennenwein} A prominent material combination is Ni$_{80}$Fe$_{20}$ (Py) and Pt bilayers. \cite{PhysRevB.85.144408, PhysRevB.82.214403} In these kind of metallic 
heterosystems, the dependencies of the spin-pumping induced ISHE voltage on the saturation magnetization and the damping constant, \cite{Yoshino2012} on the geometry \cite{PhysRevB.85.144408} and on the thicknesses of the individual layers \cite{Azevedo2011} have been investigated. Recently, it has been shown that the spin-pumping mechanism is even applicable in insulating FM/NM bilayers like  YIG/Pt (YIG:yttrium iron garnet). \cite{Kajiwara,jungfleisch:022411, Ando-nonlinear,Jungfleisch-thickness,Castel-2013,Castel,Kurebayashi,Sandweg,Chumak,Jungfleisch} Investigations on standing \cite{Jungfleisch} as well as propagating \cite{Chumak} magnons in a wide range of wavelengths \cite{Kurebayashi,Sandweg} by a combination of spin pumping and ISHE have been performed. Furthermore, the YIG and Pt thickness dependence on the ISHE-voltage from spin pumping \cite{Jungfleisch-thickness,Castel-2013,Castel} and nonlinear spin pumping \cite{Ando-nonlinear} has been observed.

\noindent 

Although spin pumping is an interfacial effect, the manner in which it, and therefore the ISHE signal strength, is affected by the structural parameters of the interface have so far not been investigated in detail. Recently, the role of the cleanliness of the YIG suface, \cite{jungfleisch:022411} and the crystal perfection of YIG  \cite{Qiu2013} in the YIG /Pt system proved the great importance of the interface properties. However in metallic systems the study of the influence of the growth modes is limited. For example, it is not clear to what extend the polycrystalline nature and the roughness of Py in metallic billayers  influences the observed SP and ISHE. In this paper, we address the problem of structural and interfacial quality by using Fe/Pt bilayers epitaxial grown on MgO substrates.  We correlate the morphology of the interface and the presence of strong anisotropy to the spin-pumping efficiency, and therefore, on the ISHE signal strength.

\begin{figure}
\includegraphics[width =0.8\columnwidth]{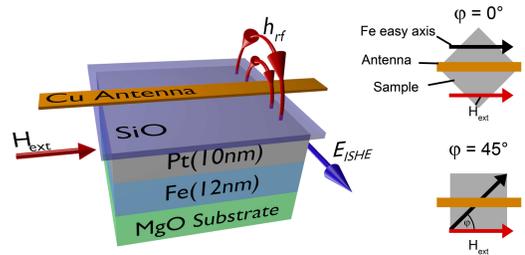}
   \caption{\label{fig_1} (left) Schematic diagram of the sample structure with the corresponding thicknesses. Ferromagnetic resonance is excited by a microwave magnetic field induced by a Cu microstrip antenna and a tunable magnetic external field (H$_\mathrm{ext}$) while a silicon oxide plate isolates the Pt from the Cu antenna. (right) Scheme of the geometry used for FMR and ISHE measurements.}
\end{figure}

\noindent Fe (12 nm)/ Pt (10 nm) bilayers were grown on MgO (100) substrates (Fig.~\ref{fig_1}) by electron-beam evaporation in an ultrahigh vacuum (UHV) chamber with a base pressure of 3 $\times$ 10$^{-11}$ mbar,  at three different substrate temperatures: room temperature ($\mathrm{RT}$), $150 ^\circ\mathrm{C}$ and $300 ^\circ\mathrm{C}$. After the deposition of Fe/Pt, annealing for 30 minutes at the corresponding growth temperature was performed. Fe (12 nm)/Al (2.5 nm) reference films were also fabricated at the same experimental conditions  as the Fe/Pt samples. The  Al  layer was deposited as a capping layer in order to form naturally an Al$_{2}$O$_{3}$ oxide protective layer.

\begin{figure}
\includegraphics[width=0.9 \columnwidth]{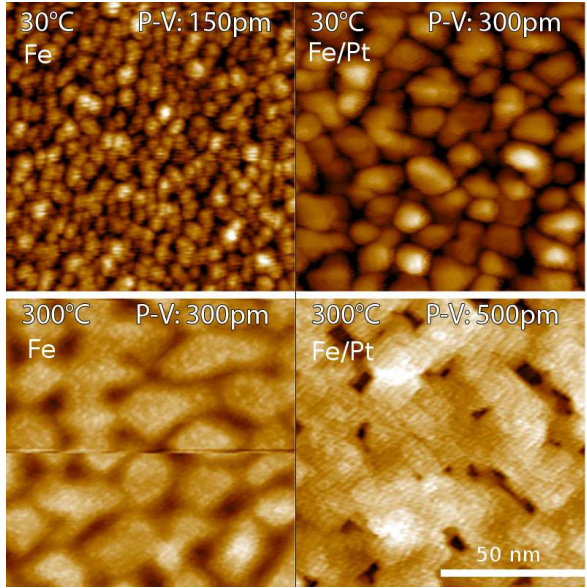}
\caption{\label{fig_stmfe} (Color online) In-situ STM images for 2 Fe/Pt samples grown at $\mathrm{RT}$ (upper panel) and $300 ^\circ\mathrm{C}$ (lower panel). The left panels show the Fe top surface after the deposition of 12 nm Fe. The right panels show the top Pt surface at the end of deposition. The color contrast corresponds to the Peak-to-Valley (P-V) values shown on the right corner of each image.}
\end{figure}

\begin{figure}
\includegraphics[width=0.9 \columnwidth]{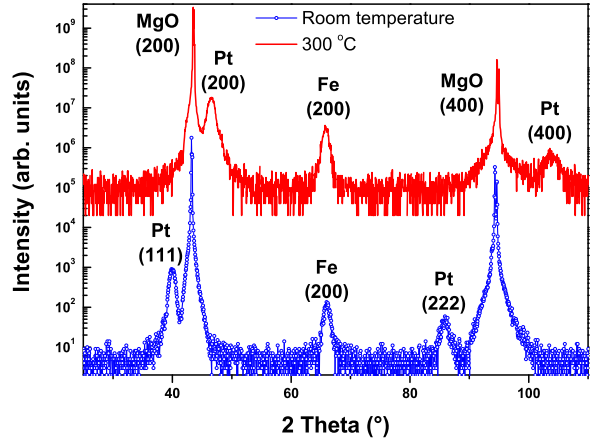}
\caption{\label{fig:xrd} (Color online) Experimental high-angle XRD patterns for two Fe/Pt samples: (bottom) grown at room temperature, and (top) grown  at $300 ^\circ\mathrm{C}$. Epitaxy of Fe on MgO is observed for all samples, while complete coherency of Pt on Fe is observed for growth above $150 ^\circ\mathrm{C}$. The curves have been shifted vertically for clarity.}
\end{figure}

\noindent In Fig.~\ref{fig_stmfe}  we present in-situ STM images for two samples grown at different temperatures: $\mathrm{RT}$ and $300 ^\circ\mathrm{C}$. The left panels show the top surface of 12 nm Fe while the right panels show the surface after 10 nm of Pt  growth. The Fe surfaces show a clear evolution with the growth temperature. The growth at  $\mathrm{RT}$ results in the creation of small crystallites with a narrow size distribution that  can be well fitted with a logarithmic - normal function, giving an average grain size of $5.94 \pm 0.79$ nm. As the growth temperature increases the average grain size become larger  being $9.5 \pm 0.71$ nm for the $150 ^\circ\mathrm{C}$ sample and $14.58 \pm 1.65$ nm for $300 ^\circ\mathrm{C}$. The size distribution is broader, and not any more well fitted by a log-norm function. Simultaneously, the mean square roughness R$^\mathrm{rms}$  increases from R$^\mathrm{rms}_{\mathrm{RT}}$ = 0.13 nm for the $\mathrm{RT}$ sample to R$^\mathrm{
rms}_{150}$ = 0.17 nm and to R$^\mathrm{rms}_{300}$ = 0.38 nm for the other two samples. While the grains themselves are essentially flat, the grain boundaries are rough leading to height differences between individual grains. Indeed, one can observe a dramatic change in the islands' shape evolving from small spheres to almost  square like crystallites for the $300 ^\circ\mathrm{C}$ samples. The grain size is strongly dependent on the ratio of the melting temperature of the deposited material (T$_\mathrm{melting}^\mathrm{element}$) to the temperature of the substrate ( T$_\mathrm{substrate}$): T$_\mathrm{melting}^\mathrm{element}$ / T$_\mathrm{substrate}$. \cite{hentzell:218,karoutsos:043525} For large values of the ratio one observes smaller grains with a narrower size distribution as for the RT sample, while for relatively smaller values of the ratio, a larger distribution of grain sizes with large grains  as seen for the other two samples. The Pt evaporation on top of Fe seems to maintain the samples' 
surface characteristics and the differences between the samples. Slightly larger grain sizes are observed. At the same time, the R$^\mathrm{rms}$ roughness is reduced for the grown at $150 ^\circ\mathrm{C}$ sample to R$^\mathrm{rms}_{150}$  = 0.13 nm and for the $300 ^\circ\mathrm{C}$ sample to R$^\mathrm{rms}_{300}$ = 0.23 nm. This is attributed to the better epitaxial relationship at the Fe/Pt interface as the X-ray diffraction (XRD) measurements reveal.

\noindent In Fig.~\ref{fig:xrd}, $\theta-2\theta$ high-angle XRD patterns for two Fe/Pt samples are presented. Epitaxy of Fe on MgO (100) is observed for all growth temperatures. For the sample grown at RT the Pt on top of Fe exhibits a preferred orientation of (111) oriented grains. In contrast, diffraction peaks of Pt (200) and Pt (400) arise from the Pt top layer for the sample grown at $300 ^\circ\mathrm{C}$ (and at $150 ^\circ\mathrm{C}$, not shown) parallel to the Fe (200) planes. The growth of the fcc Pt layer on  bcc Fe along the [100] plane direction is of great interest. This kind of epitaxy can be correlated to the Bain epitaxial orientation \cite{Daniels1994218} when the Pt cell is 45\textdegree\space rotated with respect to the Fe lattice. XRD pole figures (not shown here) reveal the complete coherency of Pt on Fe where four peaks of Pt (220) separated by 90\textdegree are rotated from the Fe (110) peaks by 45\textdegree. Furthermore, the pole figures  confirm for all samples the 45\textdegree in plane 
epitaxial relation  (Fe[100] $||$ MgO[110], Fe[110] $||$ MgO[010]) between the substrate and the Fe. The pronounced differences in the growth modes of the different Fe/Pt bilayers lead to different magnetic behaviour of the samples, different spin-pumping efficiencies and ISHE strengths, as it is shown in the next paragraph.

\begin{figure}
\includegraphics[width =0.9 \columnwidth]{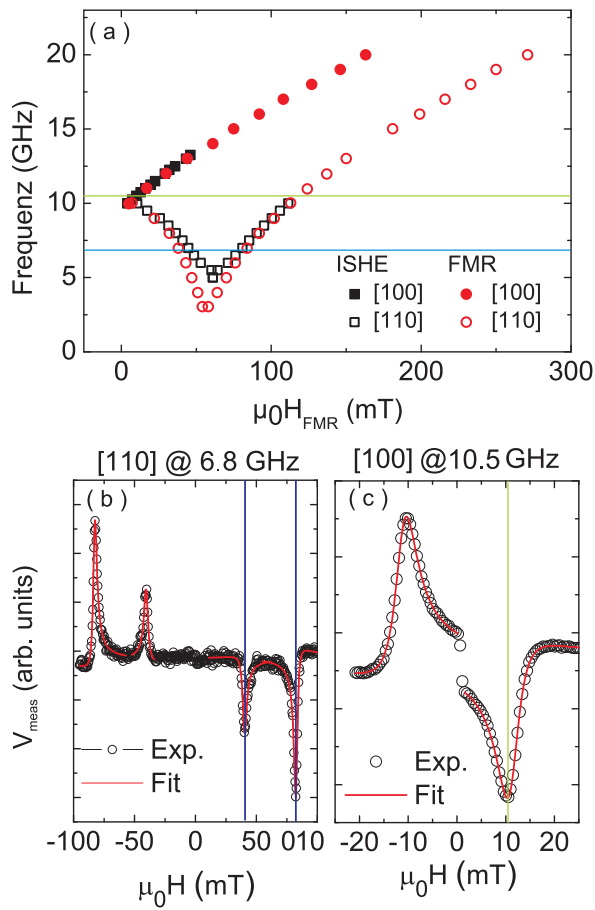}
\caption{\label{Fmr} (Color online) (a) Frequency dependence of the resonant external magnetic field along the Fe [100] easy axis direction, and along the Fe [110] hard axis direction. Two branches along [110] are observed due to pronounced anisotropy terms. (b) Measured voltage at 6.8 $\mathrm{GHz}$ along the [110] direction and (c) measured voltage at 10.5 $\mathrm{GHz}$ along [100] direction. The data refer to the sample grown at $300 ^\circ\mathrm{C}$.}
\end{figure}

\noindent In plane magnetic measurements performed with the help of the magneto-optical Kerr effect revealed for all samples the presence of a four-fold in plane magnetic anisotropy together with a uniaxial magnetic anisotropy term (UMA). The presence of this UMA reduces the cubic symmetry, making the two in-plane Fe [110] directions inequivalent. \cite{PhysRevB.44.9338} FMR and ISHE experiments were performed with the external magnetic field parallel to the film plane, in two different directions: $\varphi =0$\textdegree (Fe [100] easy axis) and $\varphi =45$\textdegree (Fe [110] hard axis) (see Fig.~\ref{fig_1}). In Fig.~\ref{Fmr}(a) the frequency dependence of the resonant field, measured in two different in plane directions and with two different setups (FMR, ISHE) are shown. The presented data refer to the sample grown at $300 ^\circ\mathrm{C}$. The form of FMR spectra along the easy and hard axis point  confirm the presence of four- and two-fold anisotropy \cite{Farle98, Zakeri2006}. 
The peak positions of the FMR signal coincide with the ISHE data. The small discrepancy for low fields can be explained by imperfect azimuthal alignment. Small variations of the in-plane angle from the hard axis can cause large shifts of the FMR peak position. \cite{Zakeri2006} 

\begin{figure}
 \includegraphics[width =1.0 \columnwidth]{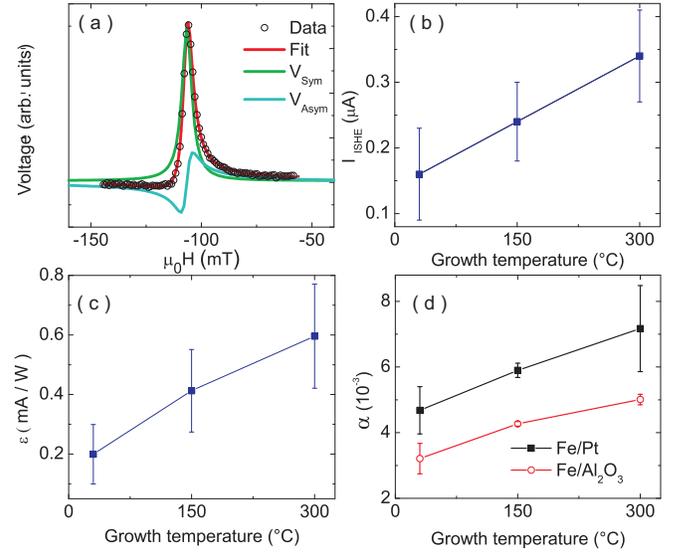}
\caption{\label{fig:isheresults} (Color online): (a) Output measured voltage form. The points refer to measured data, the red line is a fitting curve of V$_\mathrm{meas}\mathrm{(H)}$, the green and light blue lines are the symmetric and antisymmetric components obtained from the fitting, (b) ISHE current and (c) ISHE efficiency with respect to the growth temperature of Fe/Pt bilayers, (d)  difference in Gilbert damping constant for Fe reference and Fe/Pt samples. The data refer to the outer peak of ISHE voltage that corresponds to the saturated state along the hard direction (Fig.~\ref{Fmr}(b)), and were taken for an applied microwave power of 250 mW.}
\end{figure}

\noindent In FMR condition a spin current is effectively pumped from the Fe into the Pt layer. The magnetization dynamics is excited by the microwave magnetic field of a $600~\mu$m Cu microstrip antenna, and the external magnetic field is swept. We measure the spin-pumping induced ISHE-voltage \cite{jungfleisch:022411} that appears across the edges of the Pt layer perpendicular to the external magnetic field (see Fig.~\ref{fig_1} induced electromotive force E$_\mathrm{ISHE}$). In Fig.~\ref{Fmr}(b) the measured voltage (V$_\mathrm{meas}$) as a function of an external magnetic field applied in the [110] direction is presented, measured at  6.8 $\mathrm{GHz}$. With the field reversal a sign reversal in V$_\mathrm{meas}$ signal is observed at the same absolute field values. The field values where the V$_\mathrm{meas}$ peaks are measured agree with the FMR resonant positions at the 
measured frequency (denoted as horizontal lines in Fig.~\ref{Fmr}(a)) and confirm the spin pumping from Fe to Pt at the FMR frequencies. The reason for the 
appearance of two peaks is the alignment of the external field along the hard axis. Even though the external field H is increased, the intensity of the effective field  H$_\mathrm{eff}$ decreases as the direction of the magnetization is rotated from the easy axis to the hard axis. Once H$_\mathrm{eff}$ is aligned in parallel to H, H$_\mathrm{eff}$ increases. Thus, it is possible to meet the FMR condition at two different external field values. Similarly, Fig.~\ref{Fmr}(c)  shows V$_\mathrm{meas}$ with the external field H aligned along the [100] easy axis measured at a microwave frequency of 10.5 $\mathrm{GHz}$. Only one peak is observed along the easy direction as it is expected from the upper branch of the FMR spectrum of Fig.~\ref{Fmr}(a).
\noindent In both Fig.~\ref{Fmr}(b) and (c)  the amplitude and the shape of V$_\mathrm{meas}$ reveal its dependence on the magnetic anisotropy of the films. However, as it has been previously discussed, \cite{Azevedo2011,PhysRevB.82.214403,saitoh:182509} V$_\mathrm{meas}$ is the sum of three contributions: the spin pumping at the FMR frequency, the anisotropic magnetoresistance (AMR) effect at microwave fields, and the anomalous Hall effect (AHE). The possibility to distinguish between AMR and AHE contributions is given by the different dependencies of the effects with respect to the angle $\theta$ between the microwave antenna and the external magnetic field V$_\mathrm{AMR}$ $\propto \mathrm{cos(2\theta)cos(\theta)}$, V$_\mathrm{AHE}$ $\propto \mathrm{cos(\theta)}$.  \cite{Juretschke1960} Due to the strong anisotropy present in our samples, that renders the angular dependencies of V$_\mathrm{AMR}$ and V$_\mathrm{AHE}$ even more complicated, and due to limitations of our experimental setup to perform 
measurements with respect to  $\theta $ another approach has been chosen to evaluated the data. This is based on the different signal forms of symmetric and antisymmetric components of V$_\mathrm{meas}$ as it is shown in Fig.~\ref{fig:isheresults}(a). The ISHE voltage (V$_\mathrm{ISHE}$) at the FMR 
frequency obeys a Lorentz shape curve that can be decomposed into a symmetric and an antisymmetric part. \cite{Yoshino2012} 
The physical origin of the symmetric part is the ISHE itself together with contributions from AMR and AHE, while the antisymmetric part has only contributions from the AMR and AHE. The V$_\mathrm{meas}\mathrm{(H)}$ curve can be considered as a function that is formed by symmetric $V^\mathrm{Sym} (H)$  and antisymmetric $V^\mathrm{ASym}(H)$ parts, where $V^\mathrm{Sym}(H)= V_\mathrm{ISHE}(H)+V^\mathrm{Sym}_\mathrm{AMR}(H)+V^\mathrm{Sym}_\mathrm{AHE}$(H) and $V^\mathrm{ASym}(H)= V^\mathrm{ASym}_\mathrm{AMR}(H)+V^\mathrm{ASym}_\mathrm{AHE}$(H).

\noindent The red lines in Fig.~\ref{Fmr}(b) and (c) and in Fig.~\ref{fig:isheresults}(a) superimposed on the experimental black open circles data points are a result of a fit containing both symmetric and antisymmetric parts. \cite{Philipp2013} In order to separate the symmetric ISHE contribution from the overlapping symmetric AMR and AHE signals we have fabricated reference Fe/Al$_{2}$O$_{3}$ samples. By measuring the reference samples,  where no V$_\mathrm{ISHE}$ is expected,  we could calculate the symmetric parameter ratio A$^\mathrm{ref}$ by fitting the spectra of the Fe reference samples with the symmetric and antisymmetric parts,  Eq. (\ref{eq:ratio}):

\begin{equation}
\label{eq:ratio}
A^\mathrm{ref}=(V^\mathrm{Sym}_\mathrm{AMR}+V^\mathrm{Sym}_\mathrm{AHE})/(V^\mathrm{ASym}_\mathrm{AMR}+V^\mathrm{ASym}_\mathrm{AHE})
\end{equation}

\noindent The antisymmetric contributions on V$_\mathrm{meas}$ of the Fe/Pt samples can be fitted to the measured curve Fig.~\ref{Fmr}(b, c) and separated from the symmetric contribution. Assuming that the asymmetric part of the reference samples is the same as that of the Fe/Pt samples, we can estimate $V_\mathrm{ISHE}$ with the help of Eq. (\ref{eq:ratio}) by subtracting the symmetric part from the reference samples:

\begin{equation}
\label{eq:ishe_calc}
V_\mathrm{ISHE}= V^\mathrm{Sym}_\mathrm{Fe/Pt} - (V^\mathrm{Sym}_\mathrm{AMR}+V^\mathrm{Sym}_\mathrm{AHE})=V^\mathrm{Sym}_\mathrm{Fe/Pt}-A^\mathrm{ref} \cdotp V^\mathrm{ASym}_\mathrm{Fe/Pt}
\end{equation}
 
\noindent Fig.~\ref{fig:isheresults}(b, c) summarizes data referring to the outer (higher magnetic field) peak of ISHE signal along the hard direction. In Fig.~\ref{fig:isheresults}(b) the ISHE current, $I_\mathrm{ISHE}= V_\mathrm{ISHE}/R$, increases linearly with deposition temperature. Alongside the material parameters, $V_\mathrm{ISHE}$ depends also on  the absorbed microwave power at FMR frequency. To be able to compare the capability of different samples to convert an amount of absorbed energy via spin pumping into a DC current, we define the inverse spin Hall efficiency as: $\epsilon$ = I$_\mathrm{ISHE}$ / P$_\mathrm{Abs}$, where P$_\mathrm{Abs}$ is the measured absorbed microwave power via the FMR mechanism. Fig.~\ref{fig:isheresults}(c) shows that the efficiency rises with deposition temperature that provides smoother and larger grains at the interface and better quality of epitaxial samples.

\noindent Furthermore, with the help of the reference samples, we have calculated the characteristic parameter of SP, the effective spin mixing conductance g$^\mathrm{\uparrow\downarrow}$ \cite{PhysRevB.66.224403} for the Fe/Pt samples. g$^\mathrm{\uparrow\downarrow}$ is a quantifier of the efficiency of spin pumping since it describes the transfer of angular momentum at the interface to Pt. g$^\mathrm{\uparrow\downarrow}$ has been evaluated from FMR measurements from the linewidth $\Delta H$ of the FMR resonance that depends linearly on the microwave angular frequency. From the linear fit we have calculated the corresponding Gilbert damping parameter $\alpha$  for both reference Fe/Al$_{2}$O$_{3}$ and Fe/Pt samples that we plot in Fig.~\ref{fig:isheresults}(d). g$^\mathrm{\uparrow\downarrow}$ is then given by: \cite{ PhysRevB.66.224403, PhysRevB.82.214403, jungfleisch:022411}
\begin{equation}
\label{eq:spinmix}
 \mathrm{g}^\mathrm{\uparrow\downarrow} = \frac{4\pi M_\mathrm{s}^\mathrm{Fe} t_\mathrm{Fe}}{g\mu_\mathrm{B}}\Delta \alpha
\end{equation} 

\noindent where $\mathrm{\Delta \alpha} = \alpha_\mathrm{Fe/Pt}-\alpha_\mathrm{Fe/Al_{2}O_{3}}$ is the additional Gilbert damping constant due to the spin pumping. 
Eq. \ref{eq:spinmix} gives g$^\mathrm{\uparrow\downarrow}$ values that are increasing with the growth temperature and the crystalline quality of the samples. In particular, g$^\mathrm{\uparrow\downarrow}_\mathrm{RT} = (2.04 \pm 0.82)\cdotp  10^{19} \mathrm{~m^{-2}}$, g$^\mathrm{\uparrow\downarrow}_\mathrm{150^\circ\mathrm{C}} = (2.26 \pm 0.20) \cdotp 10^{19} \mathrm{~m^{-2}}$, and  g$^\mathrm{\uparrow\downarrow}_\mathrm{300^\circ\mathrm{C}} = (2.99 \pm 1.01) \cdotp 10^{19} \mathrm{~m^{-2}}$ for the samples grown at RT, $150^\circ\mathrm{C}$, and $300^\circ\mathrm{C}$ respectively. The obtained values are close to the ones observed in Py/Pt \cite{ PhysRevB.82.214403, Azevedo2011}, g$^\mathrm{\uparrow\downarrow} = (2.1 - 2.4)\cdotp 10^{19} \mathrm{m^{-2}}$  and one order of magnitude higher than in a single crystalline YIG surface in the YIG/Pt system \cite{Qiu2013} (reported values for YIG/Pt system though strongly vary in literature between $10^{16}$ m$^{-2}$ and $10^{20}$ m$^{-2}$).\cite{jungfleisch:022411}

\noindent In summary, by controlling the epitaxial relationship and the interface morphology in Fe/Pt bilayers  we were able to optimize the spin-pumping driven ISHE voltage. By increasing the growth temperature we have achieved epitaxy of Pt on Fe that led to an increase of ISHE current and ISHE efficiency. The spin-pumping induced ISHE voltage shows two distinct ISHE-voltage peaks along the hard magnetization axis revealing its strong dependence on the magnetic anisotropies. The manipulation of the magnon-to-charge current conversion with the crystal symmetry paves the way for further developments of the ISHE and its applications. 

We thank A. Taroni for helpful discussions. Financial support by Deutsche Forschungsgemeinschaft CH 1037/1-1 is gratefully acknowledged.

\begin{thebibliography}{28}%
\makeatletter
\providecommand \@ifxundefined [1]{%
 \@ifx{#1\undefined}
}%
\providecommand \@ifnum [1]{%
 \ifnum #1\expandafter \@firstoftwo
 \else \expandafter \@secondoftwo
 \fi
}%
\providecommand \@ifx [1]{%
 \ifx #1\expandafter \@firstoftwo
 \else \expandafter \@secondoftwo
 \fi
}%
\providecommand \natexlab [1]{#1}%
\providecommand \enquote  [1]{``#1''}%
\providecommand \bibnamefont  [1]{#1}%
\providecommand \bibfnamefont [1]{#1}%
\providecommand \citenamefont [1]{#1}%
\providecommand \href@noop [0]{\@secondoftwo}%
\providecommand \href [0]{\begingroup \@sanitize@url \@href}%
\providecommand \@href[1]{\@@startlink{#1}\@@href}%
\providecommand \@@href[1]{\endgroup#1\@@endlink}%
\providecommand \@sanitize@url [0]{\catcode `\\12\catcode `\$12\catcode
  `\&12\catcode `\#12\catcode `\^12\catcode `\_12\catcode `\%12\relax}%
\providecommand \@@startlink[1]{}%
\providecommand \@@endlink[0]{}%
\providecommand \url  [0]{\begingroup\@sanitize@url \@url }%
\providecommand \@url [1]{\endgroup\@href {#1}{\urlprefix }}%
\providecommand \urlprefix  [0]{URL }%
\providecommand \Eprint [0]{\href }%
\providecommand \doibase [0]{http://dx.doi.org/}%
\providecommand \selectlanguage [0]{\@gobble}%
\providecommand \bibinfo  [0]{\@secondoftwo}%
\providecommand \bibfield  [0]{\@secondoftwo}%
\providecommand \translation [1]{[#1]}%
\providecommand \BibitemOpen [0]{}%
\providecommand \bibitemStop [0]{}%
\providecommand \bibitemNoStop [0]{.\EOS\space}%
\providecommand \EOS [0]{\spacefactor3000\relax}%
\providecommand \BibitemShut  [1]{\csname bibitem#1\endcsname}%
\let\auto@bib@innerbib\@empty
\bibitem [{\citenamefont {Saitoh}\ \emph {et~al.}(2006)\citenamefont {Saitoh},
  \citenamefont {Ueda}, \citenamefont {Miyajima},\ and\ \citenamefont
  {Tatara}}]{saitoh:182509}%
  \BibitemOpen
  \bibfield  {author} {\bibinfo {author} {\bibfnamefont {E.}~\bibnamefont
  {Saitoh}}, \bibinfo {author} {\bibfnamefont {M.}~\bibnamefont {Ueda}},
  \bibinfo {author} {\bibfnamefont {H.}~\bibnamefont {Miyajima}}, \ and\
  \bibinfo {author} {\bibfnamefont {G.}~\bibnamefont {Tatara}},\ }\href
  {\doibase 10.1063/1.2199473} {\bibfield  {journal} {\bibinfo  {journal}
  {Applied Physics Letters}\ }\textbf {\bibinfo {volume} {88}},\ \bibinfo {eid}
  {182509} (\bibinfo {year} {2006})}\BibitemShut {NoStop}%
\bibitem [{\citenamefont {Kajiwara}\ \emph {et~al.}(2010)\citenamefont
  {Kajiwara}, \citenamefont {Harii}, \citenamefont {Takahashi}, \citenamefont
  {Ohe}, \citenamefont {Uchida}, \citenamefont {Mizuguchi}, \citenamefont
  {Umezawa}, \citenamefont {Kawai}, \citenamefont {Ando}, \citenamefont
  {Takanashi}, \citenamefont {Maekawa},\ and\ \citenamefont
  {Saitoh}}]{Kajiwara}%
  \BibitemOpen
  \bibfield  {author} {\bibinfo {author} {\bibfnamefont {Y.}~\bibnamefont
  {Kajiwara}}, \bibinfo {author} {\bibfnamefont {K.}~\bibnamefont {Harii}},
  \bibinfo {author} {\bibfnamefont {S.}~\bibnamefont {Takahashi}}, \bibinfo
  {author} {\bibfnamefont {J.}~\bibnamefont {Ohe}}, \bibinfo {author}
  {\bibfnamefont {K.}~\bibnamefont {Uchida}}, \bibinfo {author} {\bibfnamefont
  {M.}~\bibnamefont {Mizuguchi}}, \bibinfo {author} {\bibfnamefont
  {H.}~\bibnamefont {Umezawa}}, \bibinfo {author} {\bibfnamefont
  {H.}~\bibnamefont {Kawai}}, \bibinfo {author} {\bibfnamefont
  {K.}~\bibnamefont {Ando}}, \bibinfo {author} {\bibfnamefont {K.}~\bibnamefont
  {Takanashi}}, \bibinfo {author} {\bibfnamefont {S.}~\bibnamefont {Maekawa}},
  \ and\ \bibinfo {author} {\bibfnamefont {E.}~\bibnamefont {Saitoh}},\ }\href
  {\doibase 10.1038/nature08876} {\bibfield  {journal} {\bibinfo  {journal}
  {Nature}\ }\textbf {\bibinfo {volume} {464}},\ \bibinfo {pages} {262}
  (\bibinfo {year} {2010})}\BibitemShut {NoStop}%
\bibitem [{\citenamefont {Tserkovnyak}\ \emph
  {et~al.}(2002{\natexlab{a}})\citenamefont {Tserkovnyak}, \citenamefont
  {Brataas},\ and\ \citenamefont {Bauer}}]{Tserkovnyak}%
  \BibitemOpen
  \bibfield  {author} {\bibinfo {author} {\bibfnamefont {Y.}~\bibnamefont
  {Tserkovnyak}}, \bibinfo {author} {\bibfnamefont {A.}~\bibnamefont
  {Brataas}}, \ and\ \bibinfo {author} {\bibfnamefont {G.}~\bibnamefont
  {Bauer}},\ }\href {\doibase 10.1103/PhysRevLett.88.117601} {\bibfield
  {journal} {\bibinfo  {journal} {Physical Review Letters}\ }\textbf {\bibinfo
  {volume} {88}},\ \bibinfo {pages} {117601} (\bibinfo {year}
  {2002}{\natexlab{a}})}\BibitemShut {NoStop}%
\bibitem [{\citenamefont {Hirsch}(1999)}]{Hirsch}%
  \BibitemOpen
  \bibfield  {author} {\bibinfo {author} {\bibfnamefont {J.~E.}\ \bibnamefont
  {Hirsch}},\ }\href {\doibase 10.1103/PhysRevLett.83.1834} {\bibfield
  {journal} {\bibinfo  {journal} {Physical Review Letters}\ }\textbf {\bibinfo
  {volume} {83}},\ \bibinfo {pages} {1834} (\bibinfo {year}
  {1999})}\BibitemShut {NoStop}%
\bibitem [{\citenamefont {Czeschka}\ \emph {et~al.}(2011)\citenamefont
  {Czeschka}, \citenamefont {Dreher}, \citenamefont {Brandt}, \citenamefont
  {Weiler}, \citenamefont {Althammer}, \citenamefont {Imort}, \citenamefont
  {Reiss}, \citenamefont {Thomas}, \citenamefont {Schoch}, \citenamefont
  {Limmer}, \citenamefont {Huebl}, \citenamefont {Gross},\ and\ \citenamefont
  {Goennenwein}}]{Goennenwein}%
  \BibitemOpen
  \bibfield  {author} {\bibinfo {author} {\bibfnamefont {F.~D.}\ \bibnamefont
  {Czeschka}}, \bibinfo {author} {\bibfnamefont {L.}~\bibnamefont {Dreher}},
  \bibinfo {author} {\bibfnamefont {M.~S.}\ \bibnamefont {Brandt}}, \bibinfo
  {author} {\bibfnamefont {M.}~\bibnamefont {Weiler}}, \bibinfo {author}
  {\bibfnamefont {M.}~\bibnamefont {Althammer}}, \bibinfo {author}
  {\bibfnamefont {I.~M.}\ \bibnamefont {Imort}}, \bibinfo {author}
  {\bibfnamefont {G.}~\bibnamefont {Reiss}}, \bibinfo {author} {\bibfnamefont
  {A.}~\bibnamefont {Thomas}}, \bibinfo {author} {\bibfnamefont
  {W.}~\bibnamefont {Schoch}}, \bibinfo {author} {\bibfnamefont
  {W.}~\bibnamefont {Limmer}}, \bibinfo {author} {\bibfnamefont
  {H.}~\bibnamefont {Huebl}}, \bibinfo {author} {\bibfnamefont
  {R.}~\bibnamefont {Gross}}, \ and\ \bibinfo {author} {\bibfnamefont
  {S.~T.~B.}\ \bibnamefont {Goennenwein}},\ }\href {\doibase
  10.1103/PhysRevLett.107.046601} {\bibfield  {journal} {\bibinfo  {journal}
  {Physical Review Letters}\ }\textbf {\bibinfo {volume} {107}},\ \bibinfo
  {pages} {046601} (\bibinfo {year} {2011})}\BibitemShut {NoStop}%
\bibitem [{\citenamefont {Nakayama}\ \emph {et~al.}(2012)\citenamefont
  {Nakayama}, \citenamefont {Ando}, \citenamefont {Harii}, \citenamefont
  {Yoshino}, \citenamefont {Takahashi}, \citenamefont {Kajiwara}, \citenamefont
  {Uchida}, \citenamefont {Fujikawa},\ and\ \citenamefont
  {Saitoh}}]{PhysRevB.85.144408}%
  \BibitemOpen
  \bibfield  {author} {\bibinfo {author} {\bibfnamefont {H.}~\bibnamefont
  {Nakayama}}, \bibinfo {author} {\bibfnamefont {K.}~\bibnamefont {Ando}},
  \bibinfo {author} {\bibfnamefont {K.}~\bibnamefont {Harii}}, \bibinfo
  {author} {\bibfnamefont {T.}~\bibnamefont {Yoshino}}, \bibinfo {author}
  {\bibfnamefont {R.}~\bibnamefont {Takahashi}}, \bibinfo {author}
  {\bibfnamefont {Y.}~\bibnamefont {Kajiwara}}, \bibinfo {author}
  {\bibfnamefont {K.}~\bibnamefont {Uchida}}, \bibinfo {author} {\bibfnamefont
  {Y.}~\bibnamefont {Fujikawa}}, \ and\ \bibinfo {author} {\bibfnamefont
  {E.}~\bibnamefont {Saitoh}},\ }\href {\doibase 10.1103/PhysRevB.85.144408}
  {\bibfield  {journal} {\bibinfo  {journal} {Phys. Rev. B}\ }\textbf {\bibinfo
  {volume} {85}},\ \bibinfo {pages} {144408} (\bibinfo {year}
  {2012})}\BibitemShut {NoStop}%
\bibitem [{\citenamefont {Mosendz}\ \emph {et~al.}(2010)\citenamefont
  {Mosendz}, \citenamefont {Vlaminck}, \citenamefont {Pearson}, \citenamefont
  {Fradin}, \citenamefont {Bauer}, \citenamefont {Bader},\ and\ \citenamefont
  {Hoffmann}}]{PhysRevB.82.214403}%
  \BibitemOpen
  \bibfield  {author} {\bibinfo {author} {\bibfnamefont {O.}~\bibnamefont
  {Mosendz}}, \bibinfo {author} {\bibfnamefont {V.}~\bibnamefont {Vlaminck}},
  \bibinfo {author} {\bibfnamefont {J.~E.}\ \bibnamefont {Pearson}}, \bibinfo
  {author} {\bibfnamefont {F.~Y.}\ \bibnamefont {Fradin}}, \bibinfo {author}
  {\bibfnamefont {G.~E.~W.}\ \bibnamefont {Bauer}}, \bibinfo {author}
  {\bibfnamefont {S.~D.}\ \bibnamefont {Bader}}, \ and\ \bibinfo {author}
  {\bibfnamefont {A.}~\bibnamefont {Hoffmann}},\ }\href {\doibase
  10.1103/PhysRevB.82.214403} {\bibfield  {journal} {\bibinfo  {journal} {Phys.
  Rev. B}\ }\textbf {\bibinfo {volume} {82}},\ \bibinfo {pages} {214403}
  (\bibinfo {year} {2010})}\BibitemShut {NoStop}%
\bibitem [{\citenamefont {Yoshino}\ \emph {et~al.}(2011)\citenamefont
  {Yoshino}, \citenamefont {Ando},\ and\ \citenamefont {Harii}}]{Yoshino2012}%
  \BibitemOpen
  \bibfield  {author} {\bibinfo {author} {\bibfnamefont {T.}~\bibnamefont
  {Yoshino}}, \bibinfo {author} {\bibfnamefont {K.}~\bibnamefont {Ando}}, \
  and\ \bibinfo {author} {\bibfnamefont {K.}~\bibnamefont {Harii}},\ }\href
  {\doibase 10.1063/1.3571556} {\bibfield  {journal} {\bibinfo  {journal}
  {Applied Physics Letters}\ }\textbf {\bibinfo {volume} {132503}} (\bibinfo
  {year} {2011}),\ 10.1063/1.3571556}\BibitemShut {NoStop}%
\bibitem [{\citenamefont {Azevedo}\ \emph {et~al.}(2011)\citenamefont
  {Azevedo}, \citenamefont {Vilela-Le\~{a}o}, \citenamefont
  {Rodr\'{\i}guez-Su\'{a}rez}, \citenamefont {{Lacerda Santos}},\ and\
  \citenamefont {Rezende}}]{Azevedo2011}%
  \BibitemOpen
  \bibfield  {author} {\bibinfo {author} {\bibfnamefont {A.}~\bibnamefont
  {Azevedo}}, \bibinfo {author} {\bibfnamefont {L.~H.}\ \bibnamefont
  {Vilela-Le\~{a}o}}, \bibinfo {author} {\bibfnamefont {R.~L.}\ \bibnamefont
  {Rodr\'{\i}guez-Su\'{a}rez}}, \bibinfo {author} {\bibfnamefont {A.~F.}\
  \bibnamefont {{Lacerda Santos}}}, \ and\ \bibinfo {author} {\bibfnamefont
  {S.~M.}\ \bibnamefont {Rezende}},\ }\href {\doibase
  10.1103/PhysRevB.83.144402} {\bibfield  {journal} {\bibinfo  {journal}
  {Physical Review B}\ }\textbf {\bibinfo {volume} {83}},\ \bibinfo {pages}
  {144402} (\bibinfo {year} {2011})}\BibitemShut {NoStop}%
\bibitem [{\citenamefont {Jungfleisch}\ \emph
  {et~al.}(2013{\natexlab{a}})\citenamefont {Jungfleisch}, \citenamefont
  {Lauer}, \citenamefont {Neb}, \citenamefont {Chumak},\ and\ \citenamefont
  {Hillebrands}}]{jungfleisch:022411}%
  \BibitemOpen
  \bibfield  {author} {\bibinfo {author} {\bibfnamefont {M.~B.}\ \bibnamefont
  {Jungfleisch}}, \bibinfo {author} {\bibfnamefont {V.}~\bibnamefont {Lauer}},
  \bibinfo {author} {\bibfnamefont {R.}~\bibnamefont {Neb}}, \bibinfo {author}
  {\bibfnamefont {A.~V.}\ \bibnamefont {Chumak}}, \ and\ \bibinfo {author}
  {\bibfnamefont {B.}~\bibnamefont {Hillebrands}},\ }\href {\doibase
  10.1063/1.4813315} {\bibfield  {journal} {\bibinfo  {journal} {Applied
  Physics Letters}\ }\textbf {\bibinfo {volume} {103}},\ \bibinfo {eid}
  {022411} (\bibinfo {year} {2013}{\natexlab{a}})}\BibitemShut {NoStop}%
\bibitem [{\citenamefont {Ando}\ \emph {et~al.}(2011)\citenamefont {Ando},
  \citenamefont {An},\ and\ \citenamefont {Saitoh}}]{Ando-nonlinear}%
  \BibitemOpen
  \bibfield  {author} {\bibinfo {author} {\bibfnamefont {K.}~\bibnamefont
  {Ando}}, \bibinfo {author} {\bibfnamefont {T.}~\bibnamefont {An}}, \ and\
  \bibinfo {author} {\bibfnamefont {E.}~\bibnamefont {Saitoh}},\ }\href
  {\doibase 10.1063/1.3633348} {\bibfield  {journal} {\bibinfo  {journal}
  {Applied Physics Letters}\ }\textbf {\bibinfo {volume} {99}},\ \bibinfo
  {pages} {092510} (\bibinfo {year} {2011})}\BibitemShut {NoStop}%
\bibitem [{\citenamefont {Jungfleisch}\ \emph
  {et~al.}(2013{\natexlab{b}})\citenamefont {Jungfleisch}, \citenamefont
  {Chumak}, \citenamefont {Kehlberger}, \citenamefont {Lauer}, \citenamefont
  {Kim}, \citenamefont {Onbasli}, \citenamefont {Ross}, \citenamefont
  {Kl{\"a}ui},\ and\ \citenamefont {Hillebrands}}]{Jungfleisch-thickness}%
  \BibitemOpen
  \bibfield  {author} {\bibinfo {author} {\bibfnamefont {M.~B.}\ \bibnamefont
  {Jungfleisch}}, \bibinfo {author} {\bibfnamefont {A.~V.}\ \bibnamefont
  {Chumak}}, \bibinfo {author} {\bibfnamefont {A.}~\bibnamefont {Kehlberger}},
  \bibinfo {author} {\bibfnamefont {V.}~\bibnamefont {Lauer}}, \bibinfo
  {author} {\bibfnamefont {D.~H.}\ \bibnamefont {Kim}}, \bibinfo {author}
  {\bibfnamefont {M.~C.}\ \bibnamefont {Onbasli}}, \bibinfo {author}
  {\bibfnamefont {C.~A.}\ \bibnamefont {Ross}}, \bibinfo {author}
  {\bibfnamefont {M.}~\bibnamefont {Kl{\"a}ui}}, \ and\ \bibinfo {author}
  {\bibfnamefont {B.}~\bibnamefont {Hillebrands}},\ }\href@noop {} {\bibfield
  {journal} {\bibinfo  {journal} {arXiv:1308.3787}\ } (\bibinfo {year}
  {2013}{\natexlab{b}})}\BibitemShut {NoStop}%
\bibitem [{\citenamefont {Castel}\ \emph {et~al.}()\citenamefont {Castel},
  \citenamefont {Vlietstra}, \citenamefont {Ben~Youssef},\ and\ \citenamefont
  {van Wees}}]{Castel-2013}%
  \BibitemOpen
  \bibfield  {author} {\bibinfo {author} {\bibfnamefont {V.}~\bibnamefont
  {Castel}}, \bibinfo {author} {\bibfnamefont {N.}~\bibnamefont {Vlietstra}},
  \bibinfo {author} {\bibfnamefont {J.}~\bibnamefont {Ben~Youssef}}, \ and\
  \bibinfo {author} {\bibfnamefont {B.~J.}\ \bibnamefont {van Wees}},\
  }\href@noop {} {\bibinfo  {journal} {arXiv:1304.2190}\ }\BibitemShut
  {NoStop}%
\bibitem [{\citenamefont {Castel}\ \emph {et~al.}(2012)\citenamefont {Castel},
  \citenamefont {Vlietstra}, \citenamefont {Ben~Youssef},\ and\ \citenamefont
  {van Wees}}]{Castel}%
  \BibitemOpen
\bibfield  {journal} {  }\bibfield  {author} {\bibinfo {author} {\bibfnamefont
  {V.}~\bibnamefont {Castel}}, \bibinfo {author} {\bibfnamefont
  {N.}~\bibnamefont {Vlietstra}}, \bibinfo {author} {\bibfnamefont
  {J.}~\bibnamefont {Ben~Youssef}}, \ and\ \bibinfo {author} {\bibfnamefont
  {B.~J.}\ \bibnamefont {van Wees}},\ }\href {\doibase 10.1063/1.4754837}
  {\bibfield  {journal} {\bibinfo  {journal} {Applied Physics Letters}\
  }\textbf {\bibinfo {volume} {101}},\ \bibinfo {pages} {132414} (\bibinfo
  {year} {2012})}\BibitemShut {NoStop}%
\bibitem [{\citenamefont {Kurebayashi}\ \emph {et~al.}(2011)\citenamefont
  {Kurebayashi}, \citenamefont {Dzyapko}, \citenamefont {Demidov},
  \citenamefont {Fang}, \citenamefont {Ferguson},\ and\ \citenamefont
  {Demokritov}}]{Kurebayashi}%
  \BibitemOpen
  \bibfield  {author} {\bibinfo {author} {\bibfnamefont {H.}~\bibnamefont
  {Kurebayashi}}, \bibinfo {author} {\bibfnamefont {O.}~\bibnamefont
  {Dzyapko}}, \bibinfo {author} {\bibfnamefont {V.~E.}\ \bibnamefont
  {Demidov}}, \bibinfo {author} {\bibfnamefont {D.}~\bibnamefont {Fang}},
  \bibinfo {author} {\bibfnamefont {A.~J.}\ \bibnamefont {Ferguson}}, \ and\
  \bibinfo {author} {\bibfnamefont {S.~O.}\ \bibnamefont {Demokritov}},\ }\href
  {\doibase 10.1063/1.3652911} {\bibfield  {journal} {\bibinfo  {journal}
  {Applied Physics Letters}\ }\textbf {\bibinfo {volume} {99}},\ \bibinfo
  {pages} {162502} (\bibinfo {year} {2011})}\BibitemShut {NoStop}%
\bibitem [{\citenamefont {Sandweg}\ \emph {et~al.}(2011)\citenamefont
  {Sandweg}, \citenamefont {Kajiwara}, \citenamefont {Chumak}, \citenamefont
  {Serga}, \citenamefont {Vasyuchka}, \citenamefont {Jungfleisch},
  \citenamefont {Saitoh},\ and\ \citenamefont {Hillebrands}}]{Sandweg}%
  \BibitemOpen
  \bibfield  {author} {\bibinfo {author} {\bibfnamefont {C.~W.}\ \bibnamefont
  {Sandweg}}, \bibinfo {author} {\bibfnamefont {Y.}~\bibnamefont {Kajiwara}},
  \bibinfo {author} {\bibfnamefont {A.~V.}\ \bibnamefont {Chumak}}, \bibinfo
  {author} {\bibfnamefont {A.~A.}\ \bibnamefont {Serga}}, \bibinfo {author}
  {\bibfnamefont {V.~I.}\ \bibnamefont {Vasyuchka}}, \bibinfo {author}
  {\bibfnamefont {M.~B.}\ \bibnamefont {Jungfleisch}}, \bibinfo {author}
  {\bibfnamefont {E.}~\bibnamefont {Saitoh}}, \ and\ \bibinfo {author}
  {\bibfnamefont {B.}~\bibnamefont {Hillebrands}},\ }\href {\doibase
  10.1103/PhysRevLett.106.216601} {\bibfield  {journal} {\bibinfo  {journal}
  {Physical Review Letters}\ }\textbf {\bibinfo {volume} {106}},\ \bibinfo
  {pages} {216601} (\bibinfo {year} {2011})}\BibitemShut {NoStop}%
\bibitem [{\citenamefont {Chumak}\ \emph {et~al.}(2012)\citenamefont {Chumak},
  \citenamefont {Serga}, \citenamefont {Jungfleisch}, \citenamefont {Neb},
  \citenamefont {Bozhko}, \citenamefont {Tiberkevich},\ and\ \citenamefont
  {Hillebrands}}]{Chumak}%
  \BibitemOpen
  \bibfield  {author} {\bibinfo {author} {\bibfnamefont {A.~V.}\ \bibnamefont
  {Chumak}}, \bibinfo {author} {\bibfnamefont {A.~A.}\ \bibnamefont {Serga}},
  \bibinfo {author} {\bibfnamefont {M.~B.}\ \bibnamefont {Jungfleisch}},
  \bibinfo {author} {\bibfnamefont {R.}~\bibnamefont {Neb}}, \bibinfo {author}
  {\bibfnamefont {D.~A.}\ \bibnamefont {Bozhko}}, \bibinfo {author}
  {\bibfnamefont {V.~S.}\ \bibnamefont {Tiberkevich}}, \ and\ \bibinfo {author}
  {\bibfnamefont {B.}~\bibnamefont {Hillebrands}},\ }\href {\doibase
  10.1063/1.3689787} {\bibfield  {journal} {\bibinfo  {journal} {Applied
  Physics Letters}\ }\textbf {\bibinfo {volume} {100}},\ \bibinfo {pages}
  {082405} (\bibinfo {year} {2012})}\BibitemShut {NoStop}%
\bibitem [{\citenamefont {Jungfleisch}\ \emph {et~al.}(2011)\citenamefont
  {Jungfleisch}, \citenamefont {Chumak}, \citenamefont {Vasyuchka},
  \citenamefont {Serga}, \citenamefont {Obry}, \citenamefont {Schultheiss},
  \citenamefont {Beck}, \citenamefont {Karenowska}, \citenamefont {Saitoh},\
  and\ \citenamefont {Hillebrands}}]{Jungfleisch}%
  \BibitemOpen
  \bibfield  {author} {\bibinfo {author} {\bibfnamefont {M.~B.}\ \bibnamefont
  {Jungfleisch}}, \bibinfo {author} {\bibfnamefont {A.~V.}\ \bibnamefont
  {Chumak}}, \bibinfo {author} {\bibfnamefont {V.~I.}\ \bibnamefont
  {Vasyuchka}}, \bibinfo {author} {\bibfnamefont {A.~A.}\ \bibnamefont
  {Serga}}, \bibinfo {author} {\bibfnamefont {B.}~\bibnamefont {Obry}},
  \bibinfo {author} {\bibfnamefont {H.}~\bibnamefont {Schultheiss}}, \bibinfo
  {author} {\bibfnamefont {P.~A.}\ \bibnamefont {Beck}}, \bibinfo {author}
  {\bibfnamefont {A.~D.}\ \bibnamefont {Karenowska}}, \bibinfo {author}
  {\bibfnamefont {E.}~\bibnamefont {Saitoh}}, \ and\ \bibinfo {author}
  {\bibfnamefont {B.}~\bibnamefont {Hillebrands}},\ }\href {\doibase
  10.1063/1.3658398} {\bibfield  {journal} {\bibinfo  {journal} {Applied
  Physics Letters}\ }\textbf {\bibinfo {volume} {99}},\ \bibinfo {pages} {2512}
  (\bibinfo {year} {2011})}\BibitemShut {NoStop}%
\bibitem [{\citenamefont {Qiu}\ \emph {et~al.}(2013)\citenamefont {Qiu},
  \citenamefont {Ando},\ and\ \citenamefont {Uchida}}]{Qiu2013}%
  \BibitemOpen
  \bibfield  {author} {\bibinfo {author} {\bibfnamefont {Z.}~\bibnamefont
  {Qiu}}, \bibinfo {author} {\bibfnamefont {K.}~\bibnamefont {Ando}}, \ and\
  \bibinfo {author} {\bibfnamefont {K.}~\bibnamefont {Uchida}},\ }\href
  {http://arxiv.org/abs/1302.7091} {\bibfield  {journal} {\bibinfo  {journal}
  {arXiv:1302.7091v3}\ } (\bibinfo {year} {2013})}\BibitemShut {NoStop}%
\bibitem [{\citenamefont {Hentzell}\ \emph {et~al.}(1984)\citenamefont
  {Hentzell}, \citenamefont {Grovenor},\ and\ \citenamefont
  {Smith}}]{hentzell:218}%
  \BibitemOpen
  \bibfield  {author} {\bibinfo {author} {\bibfnamefont {H.~T.~G.}\
  \bibnamefont {Hentzell}}, \bibinfo {author} {\bibfnamefont {C.~R.~M.}\
  \bibnamefont {Grovenor}}, \ and\ \bibinfo {author} {\bibfnamefont {D.~A.}\
  \bibnamefont {Smith}},\ }\href {\doibase 10.1116/1.572727} {\bibfield
  {journal} {\bibinfo  {journal} {Journal of Vacuum Science \& Technology A:
  Vacuum, Surfaces, and Films}\ }\textbf {\bibinfo {volume} {2}},\ \bibinfo
  {pages} {218} (\bibinfo {year} {1984})}\BibitemShut {NoStop}%
\bibitem [{\citenamefont {Karoutsos}\ \emph {et~al.}(2007)\citenamefont
  {Karoutsos}, \citenamefont {Papasotiriou}, \citenamefont {Poulopoulos},
  \citenamefont {Kapaklis}, \citenamefont {Politis}, \citenamefont
  {Angelakeris}, \citenamefont {Kehagias}, \citenamefont {Flevaris},\ and\
  \citenamefont {Papaioannou}}]{karoutsos:043525}%
  \BibitemOpen
  \bibfield  {author} {\bibinfo {author} {\bibfnamefont {V.}~\bibnamefont
  {Karoutsos}}, \bibinfo {author} {\bibfnamefont {P.}~\bibnamefont
  {Papasotiriou}}, \bibinfo {author} {\bibfnamefont {P.}~\bibnamefont
  {Poulopoulos}}, \bibinfo {author} {\bibfnamefont {V.}~\bibnamefont
  {Kapaklis}}, \bibinfo {author} {\bibfnamefont {C.}~\bibnamefont {Politis}},
  \bibinfo {author} {\bibfnamefont {M.}~\bibnamefont {Angelakeris}}, \bibinfo
  {author} {\bibfnamefont {T.}~\bibnamefont {Kehagias}}, \bibinfo {author}
  {\bibfnamefont {N.~K.}\ \bibnamefont {Flevaris}}, \ and\ \bibinfo {author}
  {\bibfnamefont {E.~T.}\ \bibnamefont {Papaioannou}},\ }\href {\doibase
  10.1063/1.2769785} {\bibfield  {journal} {\bibinfo  {journal} {Journal of
  Applied Physics}\ }\textbf {\bibinfo {volume} {102}},\ \bibinfo {eid}
  {043525} (\bibinfo {year} {2007})}\BibitemShut {NoStop}%
\bibitem [{\citenamefont {Daniels}\ \emph {et~al.}(1994)\citenamefont
  {Daniels}, \citenamefont {Nix},\ and\ \citenamefont
  {Clemens}}]{Daniels1994218}%
  \BibitemOpen
  \bibfield  {author} {\bibinfo {author} {\bibfnamefont {B.}~\bibnamefont
  {Daniels}}, \bibinfo {author} {\bibfnamefont {W.}~\bibnamefont {Nix}}, \ and\
  \bibinfo {author} {\bibfnamefont {B.}~\bibnamefont {Clemens}},\ }\href
  {\doibase http://dx.doi.org/10.1016/0040-6090(94)90323-9} {\bibfield
  {journal} {\bibinfo  {journal} {Thin Solid Films}\ }\textbf {\bibinfo
  {volume} {253}},\ \bibinfo {pages} {218 } (\bibinfo {year}
  {1994})}\BibitemShut {NoStop}%
\bibitem [{\citenamefont {Florczak}\ and\ \citenamefont
  {Dahlberg}(1991)}]{PhysRevB.44.9338}%
  \BibitemOpen
  \bibfield  {author} {\bibinfo {author} {\bibfnamefont {J.~M.}\ \bibnamefont
  {Florczak}}\ and\ \bibinfo {author} {\bibfnamefont {E.~D.}\ \bibnamefont
  {Dahlberg}},\ }\href {\doibase 10.1103/PhysRevB.44.9338} {\bibfield
  {journal} {\bibinfo  {journal} {Phys. Rev. B}\ }\textbf {\bibinfo {volume}
  {44}},\ \bibinfo {pages} {9338} (\bibinfo {year} {1991})}\BibitemShut
  {NoStop}%
\bibitem [{\citenamefont {Farle}(1998)}]{Farle98}%
  \BibitemOpen
  \bibfield  {author} {\bibinfo {author} {\bibfnamefont {M.}~\bibnamefont
  {Farle}},\ }\href {http://stacks.iop.org/0034-4885/61/i=7/a=001} {\bibfield
  {journal} {\bibinfo  {journal} {Reports on Progress in Physics}\ }\textbf
  {\bibinfo {volume} {61}},\ \bibinfo {pages} {755} (\bibinfo {year}
  {1998})}\BibitemShut {NoStop}%
\bibitem [{\citenamefont {Zakeri}\ \emph {et~al.}(2006)\citenamefont {Zakeri},
  \citenamefont {Kebe}, \citenamefont {Lindner},\ and\ \citenamefont
  {Farle}}]{Zakeri2006}%
  \BibitemOpen
  \bibfield  {author} {\bibinfo {author} {\bibfnamefont {K.}~\bibnamefont
  {Zakeri}}, \bibinfo {author} {\bibfnamefont {T.}~\bibnamefont {Kebe}},
  \bibinfo {author} {\bibfnamefont {J.}~\bibnamefont {Lindner}}, \ and\
  \bibinfo {author} {\bibfnamefont {M.}~\bibnamefont {Farle}},\ }\href
  {\doibase 10.1016/j.jmmm} {\bibfield  {journal} {\bibinfo  {journal} {Journal
  of Magnetism and Magnetic Materials}\ }\textbf {\bibinfo {volume} {299}},\
  \bibinfo {pages} {L1 } (\bibinfo {year} {2006})}\BibitemShut {NoStop}%
\bibitem [{\citenamefont {Juretschke}(1960)}]{Juretschke1960}%
  \BibitemOpen
  \bibfield  {author} {\bibinfo {author} {\bibfnamefont {H.~J.}\ \bibnamefont
  {Juretschke}},\ }\href {\doibase 10.1063/1.1735851} {\bibfield  {journal}
  {\bibinfo  {journal} {Journal of Applied Physics}\ }\textbf {\bibinfo
  {volume} {31}},\ \bibinfo {pages} {1401} (\bibinfo {year}
  {1960})}\BibitemShut {NoStop}%
\bibitem [{\citenamefont {Fuhrmann}(2013)}]{Philipp2013}%
  \BibitemOpen
  \bibfield  {author} {\bibinfo {author} {\bibfnamefont {P.}~\bibnamefont
  {Fuhrmann}},\ }\href@noop {} {\bibfield  {journal} {\bibinfo  {journal}
  {Diplomarbeit, Technische Universit\"{a}t Kaiserslautern}\ } (\bibinfo {year}
  {2013})}\BibitemShut {NoStop}%
\bibitem [{\citenamefont {Tserkovnyak}\ \emph
  {et~al.}(2002{\natexlab{b}})\citenamefont {Tserkovnyak}, \citenamefont
  {Brataas},\ and\ \citenamefont {Bauer}}]{PhysRevB.66.224403}%
  \BibitemOpen
  \bibfield  {author} {\bibinfo {author} {\bibfnamefont {Y.}~\bibnamefont
  {Tserkovnyak}}, \bibinfo {author} {\bibfnamefont {A.}~\bibnamefont
  {Brataas}}, \ and\ \bibinfo {author} {\bibfnamefont {G.~E.~W.}\ \bibnamefont
  {Bauer}},\ }\href {\doibase 10.1103/PhysRevB.66.224403} {\bibfield  {journal}
  {\bibinfo  {journal} {Phys. Rev. B}\ }\textbf {\bibinfo {volume} {66}},\
  \bibinfo {pages} {224403} (\bibinfo {year} {2002}{\natexlab{b}})}\BibitemShut
  {NoStop}%
\end{thebibliography}

%

\end{document}